\documentclass[12pt]{article}
\begin{document}
\title{THE GAUGE HIERARCHY PROBLEM AND PLANCK OSCILLATORS}
\author{B.G. Sidharth\\
International Institute for Applicable Mathematics \& Information Sciences\\
Hyderabad (India) \& Udine (Italy)\\
B.M. Birla Science Centre, Adarsh Nagar, Hyderabad - 500 063 (India)}
\date{}
\maketitle
\begin{abstract}
A longstanding question that has puzzled Physicists is the so called gauge hierarchy problem, that is why is there such a wide gap between the mass of a Planck particle, $10^{-5}gms$ and the mass of a typical elementary particle $\sim 10^{-25}gms$. We show that the answer to this problem lies in a particular characterization of gravitation. This moreover also provides a picture of a Planck scale underpinning for the entire universe itself.
\end{abstract}
\section{Introduction}
In 1997 the author put forward a model in which particles are created fluctuationally in a phase transition from the background Zero Point Field or Dark Energy. This model lead to dramatic consequences: The Universe would be accelerating and expanding with a small cosmological constant. Besides, several longstanding puzzling relations of the so called large number genre which had no explanation whatsoever, were now deduceable from the theory.\\
At that time the accepted Standard Big Bang model predicted a dark matter filled decelerating universe--in other words the exact opposite. In 1998 however the first results were announced by Perlmutter, Kirshner and coworkers, from a study of distant supernovae that the universe was indeed accelerating and expanding with a small cosmological constant, contrary to belief. These conclusions were subsequently reconfirmed several times. This work was the breakthrough of the year 1998 of the American Association for the Advancement of Science \cite{science} while Dark Energy itself was subsequently confirmed through the WMAP and the Sloan Digital Sky Survey. This in fact was the breakthrough of the year 2003 \cite{science2}.\\
To recapitulate the author's model \cite{mg8,ijmpa,ijtp,csfnc,csfnc2,cu,uof}, we give a simple picture and will return to the nuances later.\\
Our starting point is the all permeating Zero Point Field or the Dark Energy, from which the elementary particles are created. As Wheeler put it \cite{mwt}, ``From the zero-point fluctuations of a single oscillator to the fluctuations of the electromagnetic field to geometrodynamic fluctuations is a natural 
order of progression...''\\ Let us consider,
following Wheeler a harmonic oscillator in its ground state. The
probability amplitude is
$$\psi (x) = \left(\frac{m\omega}{\pi \hbar}\right)^{1/4}
e^{-(m\omega/2\hbar)x^2}$$ for displacement by the distance $x$ from
its position of classical equilibrium. So the oscillator fluctuates
over an interval
$$\Delta x \sim (\hbar/m\omega)^{1/2}$$ The electromagnetic field is
an infinite collection of independent oscillators, with amplitudes
$X_1,X_2$ etc. The probability for the various oscillators to have
emplitudes $X_1, X_2$ and so on is the product of individual
oscillator amplitudes:
$$\psi (X_1,X_2,\cdots ) = exp [-(X^2_1 + X^2_2 + \cdots)]$$ wherein
there would be a suitable normalization factor. This expression gives
the probability amplitude $\psi$ for a configuration $B (x,y,z)$ of
the magnetic field that is described by the Fourier coefficients
$X_1,X_2,\cdots$ or directly in terms of the magnetic field
configuration itself by
$$\psi (B(x,y,z)) = P exp \left(-\int \int \frac{\bf{B}(x_1)\cdot
\bf{B}(x_2)}{16\pi^3\hbar cr^2_{12}} d^3x_1 d^3x_2\right).$$ $P$ being
a normalization factor. Let us consider a configuration where the
magnetic field is everywhere zero except in a region of dimension $l$,
where it is of the order of $\sim \Delta B$. The probability amplitude
for this configuration would be proportional to
$$\exp [-(\Delta B)^2 l^4/\hbar c)$$ So the energy
 of fluctuation in
a region of length $l$ is given by finally \cite{mwt,r24,r25}
$$B^2 \sim \frac{\hbar c}{l^4}$$
 In the above if $l$ is taken to be
the Compton wavelength of a typical elementary
 particle, then we
recover its energy $mc^2$, as can be easily verified.\\
It may
 be mentioned
that Einstein himself had believed that the electron was a result
 of
such a condensation from the background electromagnetic field
(Cf.\cite{r26,cu}
 for details). We will return to this point
again. We also take the pion to represent a typical elementary
particle, as in the literature.\\
 
To proceed, as there are $N \sim
10^{80}$ such particles in the universe,
 we get
\begin{equation}
Nm = M\label{e1}
\end{equation}
where $M$ is the mass of the universe.\\
 
In the following we will
use $N$ as the sole cosmological parameter.\\
 
Equating the
gravitational potential energy of the pion in a three dimensional
isotropic
 sphere of pions of radius $R$, the radius of the universe,
with the rest
 energy of the pion, we can deduce the well known
relation \cite{r27,r28,r29}
\begin{equation}
R \approx \frac{GM}{c^2}\label{e2}
\end{equation}
where $M$ can be obtained from (\ref{e1}).\\
 
We now use the fact
that given $N$ particles, the fluctuation in the particle number is of
the
 order $\sqrt{N}$\cite{r29,r30,ijmpa,ijtp,r5,mg8}, while a typical time
interval for the
 fluctuations is $\sim \hbar/mc^2$, the Compton
time, the fuzzy interval we encountered in the previous Chapter.  We
will come back to
 this point later in this Chapter, in the context
of the minimum Planck scale: Particles are created and destroyed - but
the ultimate result is that $\sqrt{N}$ particles are created. So we
have, as we saw briefly earlier,
\begin{equation}
\frac{dN}{dt} = \frac{\sqrt{N}}{\tau}\label{ex}
\end{equation}
whence on integration we get, (remembering that we are almost in the
continuum region),
\begin{equation}
T = \frac{\hbar}{mc^2} \sqrt{N}\label{e3}
\end{equation}
Later we will analyze in a little greater detail, the above and subsequent relations and continue with
this preliminary treatment.  We can easily verify that the equation is
indeed satisfied
 where $T$ is the age of the universe. Next by
differentiating (\ref{e2}) with
 respect to $t$ we get
\begin{equation}
\frac{dR}{dt} \approx HR\label{e4}
\end{equation}
where $H$ in (\ref{e4}) can be identified with the Hubble Constant,
and using
 (\ref{e2}) is given by,
\begin{equation}
H = \frac{Gm^3c}{\hbar^2}\label{e5}
\end{equation}
Equation (\ref{e1}), (\ref{e2}) and (\ref{e3}) show that in this
formulation, the correct mass,
 radius, Hubble constant and age of
the universe can be deduced given $N$ as the sole
 cosmological or
large scale parameter. Equation (\ref{e5}) can be written as
\begin{equation}
m \approx \left(\frac{H\hbar^2}{Gc}\right)^{\frac{1}{3}}\label{e6}
\end{equation}
Equation (\ref{e6}) has been empirically known as an "accidental" or
"mysterious" relation.
 As observed by Weinberg\cite{r10}, this is
unexplained: it relates a single
 cosmological parameter $H$ to
constants from microphysics. We will touch upon
 this micro-macro
nexus again.
 In our formulation, equation (\ref{e6}) is no longer a
mysterious coincidence but
 rather a consequence.\\
 
As (\ref{e5})
and (\ref{e4}) are not exact equations but rather,
 order of
magnitude relations, it follows, on differentiating (\ref{e4}) that
a
 small cosmological constant $\wedge$ is allowed such that
$$\wedge < 0 (H^2)$$
 This is consistent with observation and shows
that $\wedge$ is very small -- this has been a puzzle,
 the so called
cosmological constant problem alluded to, because in conventional
theory, it turns out to be huge \cite{rj}. Some $10^{70}$ times higher in fact! But it poses no problem in
this formulation.\\
 
To proceed we observe that because of the fluctuation of
$\sim \sqrt{N}$ (due to the ZPF),
 there is an excess electrical
potential energy of the electron, which infact
 we have identified as
its inertial energy. That is \cite{ijmpa,r29},
$$\sqrt{N} e^2/R \approx mc^2.$$
 On using (\ref{e2}) in the above,
we recover the well known Gravitation-electromagnetism
 ratio viz.,
\begin{equation}
e^2/Gm^2 \sim \sqrt{N} \approx 10^{40}\label{e7}
\end{equation}
or without using (\ref{e2}), we get, instead, the well known so
called
 Weyl-Eddington formula,
\begin{equation}
R = \sqrt{N}l\label{e8}
\end{equation}
(It appears that this was first noticed by H. Weyl \cite{r31}).
Infact (\ref{e8}) is the spatial counterpart of (\ref{e3}). If we
combine (\ref{e8}) and (\ref{e2}), we get,
\begin{equation}
\frac{Gm}{lc^2} = \frac{1}{\sqrt{N}} \propto T^{-1}\label{e9}
\end{equation}
where in (\ref{e9}), we have used (\ref{e3}). This was the relation we
encountered in the previous chapter. Following Dirac (cf.also
\cite{r32})
 we treat $G$ as the variable, rather than the quantities
$m, l, c \,\mbox{and}\,
 \hbar$ (which we will call micro physical
constants) because of their central role
 in atomic (and sub atomic)
physics.\\
 Next if we use $G$ from (\ref{e9}) in (\ref{e5}), we can
see that
\begin{equation}
H = \frac{c}{l} \quad \frac{1}{\sqrt{N}}\label{e10}
\end{equation}
Thus apart from the fact that $H$ has the same inverse time dependance
on
 $T$ as $G$, (\ref{e10}) shows that given the microphysical
constants, and
 $N$, we can deduce the Hubble Constant also, as from
(\ref{e10}) or (\ref{e5}).\\
 Using (\ref{e1}) and (\ref{e2}), we can
now deduce that
\begin{equation}
\rho \approx \frac{m}{l^3} \quad \frac{1}{\sqrt{N}}\label{e11}
\end{equation}
Next (\ref{e8}) and (\ref{e3}) give,
\begin{equation}
R = cT\label{e12}
\end{equation}
(\ref{e11}) and (\ref{e12}) are consistent with observation.\\
With regard to the time variation of $G$, the issue is debatable and model dependent. Measurements on the earth and of the planets, and perhaps most accurate of all, Pulsars indicates a value $\sim 10^{-10}$, though values $10^{-11}$ and $10^{-12}$ have also appeared in some studies \cite{mel,uzan}.
\section{The Gauge Hierarchy Problem}
A long standing puzzle has been the so called Gauge Hierarchy Problem. This deals with the fact that the Planck mass is some $10^{20}$ times the mass of an elementary particle, for example Gauge Bosons or Protons or Electrons (in the large number sense). Why is there such a huge gap?\\
It is well known that the Planck mass is defined by \cite{mwt,cu}
\begin{equation}
m_P = \left(\frac{\hbar c}{G}\right)^{1/2} \, \sim 10^{-5}gm\label{e1a}
\end{equation}
Alternatively the Planck mass defines a black hole having the Schwarzchild radius given by
\begin{equation}
\frac{2Gm_P}{c^2} \sim l_P \, \sim 10^{-33}cm\label{e2a}
\end{equation}
In (\ref{e2}) $l_P$ is the Planck length. Interestingly Rosen \cite{rosen} has shown that the Planck mass given in (\ref{e1a}) is an universe in itself, that is a black hole given by (\ref{e2a}).\\
While $m_P$ is $\sim 10^{-5}gms$, a typical elementary particle has a mass $m \sim 10^{-25}gms$. (As mentioned in this order of magnitude sense it does not make much difference, if the elementary particle is an electron or pion or proton (Cf.ref.\cite{mwt})).\\
We now recall that as already shown we have \cite{ijmpa,ijtp,fpl,uof}
\begin{equation}
G = \frac{\hbar c}{m^2\sqrt{N}}\label{e3a}
\end{equation}
In (\ref{e3a}) $N \sim 10^{80}$ is the well known number of elementary particles in the universe, which features in the Weyl-Eddington relations as also the Dirac Cosmology.\\
What is interesting about (\ref{e3a}) is that it shows gravitation as a distributional effect over all the $N$ particles in the universe \cite{fpl,uof}.\\
Let us rewrite (\ref{e2a}) in the form
\begin{equation}
G \approx \frac{\hbar c}{m^2_P}\label{e4a}
\end{equation}
remembering that the Planck length is also the Compton length of the Planck mass. (Interestingly an equation like (\ref{e2a}) or (\ref{e4a}) also follows from Sakharov's treatment of gravitation \cite{sakharov}.) A division of (\ref{e3}) and (\ref{e4a}) yields
\begin{equation}
m^2_P = \sqrt{N} m^2\label{e5a}
\end{equation}
Equation (\ref{e5a}) immediately gives the ratio $\sim 10^{20}$ between the Planck mass and the mass of an elementary particle.\\
It is interesting that in (\ref{e3a}) if we take $N \sim 1$, then we recover (\ref{e4a}). So while the Planck mass in the spirit of Rosen's isolated universe and the Schwarzchild black hole uses the gravitational interaction in isolation, as seen from (\ref{e3a}), elementary particles are involved in the gravitational interaction with all the remaining particles in the universe.\\
Finally rememebring that $Gm_P^2 \sim e^2$, as can also be seen from (\ref{e4}), we get from (\ref{e3a})
\begin{equation}
\frac{e^2}{Gm^2} \sim \frac{1}{\sqrt{\bar{N}}}\label{e6a}
\end{equation}
Equation (\ref{e6a}) is the otherwise empirically well known electromagnetism-gravitation coupling constant ratio, but here it is deduced from the theory.\\
It may be remarked that one could attempt an explanation of (\ref{e5a}) from the point of view of SuperSymmetry or Brane theory, but these latter have as yet no experimental validation \cite{gor}.
\section{The Planck Scale}
It is now generally accepted in various Quantum Gravity approaches as also Quantum Super String Theory or M-Theory that the universe has a minimum scale--the Planck scale. It is well known, and this was realized by Planck himself that there is an absolute minimum scale in the universe, and this is,
$$l_P = \left(\frac{\hbar G}{c^3}\right)^{\frac{1}{2}} \sim
10^{-33}cm$$
\begin{equation}
t_P = \left(\frac{\hbar G}{c^5}\right)^{\frac{1}{2}} \sim
10^{-42}sec\label{ea1}
\end{equation}
Yet what we encounter in the real world is, not the Planck scale,
but the elementary particle Compton scale. The explanation given for
this is that the very high energy Planck scale is moderated by the
Uncertainty Principle. The question which arises is, exactly how
does this happen? We will now present an argument to show how the
Planck scale leads to the real world Compton scale, via
fluctuations and the modification of the Uncertainty Principle.\\
We note that (\ref{ea1}) defines the absolute minimum physical scale
\cite{bgsafdb,mwt,rr2,garay}. Associated with (\ref{ea1}) is the Planck mass
\begin{equation}
m_P \sim 10^{-5}gm\label{ea2}
\end{equation}
There are certain interesting properties associated with
(\ref{ea1}) and (\ref{ea2}). $l_P$ is the Schwarzschild radius of a
black hole of mass $m_P$ while $t_P$ is the evaporation time for
such a black hole via the Beckenstein radiation \cite{rr3}.
Interestingly $t_P$ is also the Compton time for the Planck mass,
a circumstance that is symptomatic of the fact that at this scale,
electromagnetism and gravitation become of the same order
\cite{cu}. Indeed all this fits in very well with Rosen's analysis
that such a Planck scale particle would be a mini universe
\cite{rosen,rr6}. We will now re-invoke a time varying gravitational
constant
\begin{equation}
G \approx \frac{lc^2}{m\sqrt{N}} \propto (\sqrt{N}t)^{-1} \propto
T^{-1}\label{ea3}
\end{equation}
In (\ref{ea3}) $m$ and $l$ are the
mass and Compton wavelength of a typical elementary particle like
the pion while $N \sim 10^{80}$ is the number of elementary
particles in the universe, and
$T$ is the age of the universe.\\
So, 
$$\frac{dN}{dt} = \frac{\sqrt{N}}{\tau}$$
whence on integration we get,
$$T = \frac{\hbar}{mc^2} \sqrt{N}$$
and we can also deduce its spatial counterpart, $R = \sqrt{N} l$,
which is the well known empirical Eddington formula.\\
Equation (\ref{ea3}) which is an order of magnitude relation is
consistent with observation as noted \cite{rr11,mel} while it may be remarked
that the Dirac cosmology itself has inconsistencies.\\
Substitution of (\ref{ea3}) in (\ref{ea1}) yields
$$l = N^{\frac{1}{4}} l_P,$$
\begin{equation}
t = N^{\frac{1}{4}} t_P\label{ea4}
\end{equation}
where $t$ as noted is the typical Compton time of an elementary
particle. We can easily verify that (\ref{ea4}) is correct. It
must be stressed that (\ref{ea4}) is not a fortuitous empirical
coincidence, but rather is a result of using (\ref{ea3}), which
again as noted, can be deduced from
theory.\\
(\ref{ea4}) can be rewritten as
$$l = \sqrt{n}l_P$$
\begin{equation}
t = \sqrt{n}t_P\label{ea5}
\end{equation}
wherein we have used (\ref{ea1}) and (\ref{ea3}) and $n = \sqrt{N}$.\\
We will now compare (\ref{ea5}) with the well known relations, referred to earlier,
\begin{equation}
R = \sqrt{N} l \quad T = \sqrt{N} t\label{ea6}
\end{equation}
The first relation of (\ref{ea6}) is the well known Weyl-Eddington
formula referred to while the second relation of (\ref{ea6}) is
given also on the right side of (\ref{ea3}). We now observe that
(\ref{ea6}) can be seen to be the result of a Brownian Walk
process, $l,t$ being typical intervals between "steps"
(Cf.\cite{cu,rr12,rr13}). We demonstrate this below after equation
(\ref{ea8}). On the other hand, the typical intervals $l,t$ can be
seen to result from a diffusion  process themselves. Let us
consider the well known diffusion relation,
\begin{equation}
(\Delta x)^2 \equiv l^2 = \frac{\hbar}{m} t \equiv \frac{\hbar}{m}
\Delta t\label{ea7}
\end{equation}
(Cf.\cite{rr12},\cite{rr14}-\cite{rr17}). What is being done here is that we are modeling fuzzy spacetime by a double Wiener process to be touched upon later, which leads to (\ref{ea7}). This will be seen in more detail, below.\\
Indeed as $l$ is the Compton wavelength, (\ref{ea7}) can be
rewritten as the Quantum Mechanical Uncertainty Principle
$$l \cdot p \sim \hbar$$
at the Compton scale (Cf. also \cite{rr18}) (or even at the de
Broglie
scale).\\
What (\ref{ea7}) shows is that a Brownian process defines
the Compton scale while (\ref{ea6}) shows that a Random Walk
process with the Compton scale as the interval defines the length
and time scales of the universe itself (Cf.\cite{rr13}). Returning
now to (\ref{ea5}), on using (\ref{ea2}), we observe that in
complete analogy with (\ref{ea7}) we have the relation
\begin{equation}
(\Delta x)^2 \equiv l^2_P = \frac{\hbar}{m_P} t_P \equiv
\frac{\hbar}{m_P} \Delta t\label{ea8}
\end{equation}
We can now argue that the Brownian process (\ref{ea8})
defines the Planck length while a Brownian Random Walk process
with the Planck scale as the interval leads to (\ref{ea5}), that is
the
Compton scale.\\
To see all this in greater detail, it may be observed that
equation (\ref{ea8}) (without subscripts)
\begin{equation}
(\Delta x)^2 = \frac{\hbar}{m} \Delta t\label{eaa}
\end{equation}
is the same as the equation (\ref{ea7}), indicative of a double
Wiener process. Indeed as noted by several scholars, this defines
the fractal Quantum path of
dimension 2 (rather than dimension 1) (Cf.e.g. ref.\cite{rr15}).\\
Firstly it must be pointed out that equation (\ref{eaa}) defines a
minimum space time unit - the Compton scale $(l,t)$. This follows
from (\ref{eaa}) if we substitute into it $\langle \frac{\Delta x}
{\Delta t}\rangle_{max} = c$. If the mass of the particle is the
Planck mass, then this Compton scale becomes the Planck scale.\\
Let us now consider the distance traversed by a particle with the
speed of light through the time interval $T$. The distance $R$
covered would be
\begin{equation}
\int dx = R = c \int dt = cT\label{eIa}
\end{equation}
by conventional reasoning. In view of the equation
(\ref{eaa}), however we would have to consider firstly, the minimum
time interval $t$ (Compton or Planck time), so that we have
\begin{equation}
\int dt \to nt\label{eIIa}
\end{equation}
Secondly, because the square of the space interval $\Delta x$
(rather than the interval $\Delta x$ itself as in conventional
theory) appears in (\ref{eaa}), the left side of (\ref{eIa})
becomes, on using (\ref{eIIa})
\begin{equation}
\int dx^2 \to \int (\sqrt{n}dx) (\sqrt{n}dy)\label{eIIIa}
\end{equation}
Whence for the linear dimension $R$ we would have
\begin{equation}
\sqrt{n}R = nct \quad \mbox{or} \quad R = \sqrt{n} l\label{eIVa}
\end{equation}
Equation (\ref{eIIIa}) brings out precisely the fractal dimension
$D = 2$ of the Brownian path while (\ref{eIVa}) is identical to
(\ref{ea4}) or (\ref{ea6}) (depending on whether we are dealing with
minimum intervals of the Planck scale or Compton scale of
elementary particles). Apart from showing the Brownian character
linking equations (\ref{ea4}) and (\ref{eaa}), incidentally, this
also provides the justification for what has so far been
considered to be a mysterious large number coincidence viz. the
Eddington
formula (\ref{ea6}).\\
There is another way of looking at this. It is well known that in approaches like that of the author or 
Quantum Super String Theory, at the Planck scale we have a non
commutative geometry \cite{rr19,bgsgrav}
Indeed as noted, (\ref{e3}) follows without recourse to Quantum
Super Strings, merely by the fact that $l_P,t_P$ are the absolute
minimum space
time intervals as we saw earlier.\\
The non commutative geometry (\ref{e3}), as is known, is
symptomatic of a modified uncertainty principle at this scale
\cite{rr22}-\cite{rr28}
\begin{equation}
\Delta x \approx \frac{\hbar}{\Delta p} + l^2_P \frac{\Delta
p}{\hbar}\label{ea10}
\end{equation}
The relation (\ref{ea10}) would be true even in Quantum Gravity.
The extra or second term on the right side of (\ref{ea10}) as noted in Chapter 3 expresses the well known duality effect - as we attempt to go down
to the Planck scale, infact we are lead to the larger scale. The question is, what is this larger scale? If
we now use the fact that $\sqrt{n}$ is the fluctuation in the
number of Planck particles (exactly as $\sqrt{N}$ was the
fluctuation in the particle number as in (\ref{ea3})) so that
$\sqrt{n}mpc = \Delta p$ is the fluctuation or uncertainty in the
momentum for the second term on the right side of (\ref{ea10}), we
obtain for the uncertainty in length,
\begin{equation}
\Delta x = l^2_P \frac{\sqrt{n}m_Pc}{\hbar} =
l_P\sqrt{n},\label{ea11}
\end{equation}
We can easily see that (\ref{ea11}) is the same as the first
relation of (\ref{ea5}). The second relation of (\ref{ea5}) follows
from an
application of the time analogue of (\ref{ea10}).\\
Thus the impossibility of going down to the Planck scale because
of (\ref{e3}) or (\ref{ea10}), manifests itself in the fact that as
we attempt to go down to the Planck scale, we infact end up at the
Compton scale. In the next section we will give another demonstration of this result. This is how the Compton scale is encountered in real life.\\
Interestingly while at the Planck length, we have a life time of
the order of the Planck time, as noted above it is possible to
argue on the other hand that with the  mass and length of a
typical elementary particle like the pion, at the Compton scale,
we have a life time which is the age of the universe itself as
shown by
Sivaram \cite{rr3,rr29}.\\
Interestingly also Ng and Van Dam deduce the relations like
\cite{rr30}
\begin{equation}
\delta L \leq (Ll^2_P)^{1/3}, \delta T \leq
(Tt^2_P)^{1/3}\label{ea9}
\end{equation}
where the left side of (\ref{ea9}) represents the uncertainty in the measurement
of length and time for an interval $L,T$. We would like to point
out that if in the above we use for $L,T$, the size and age of
the universe, then $\Delta L$ and
$\Delta T$ reduce to the Compton scale $l,t$.\\
In conclusion, Brownian double Wiener processes and the modification
of the Uncertainity Principle at the Planck scale lead to the
physical Compton scale.
\section{The Universe as Planck Oscillators}
In the previous section, we had argued that a typical
elementary particle like a pion could be considered to be the result of
$n \sim 10^{40}$ evanescent Planck scale particles. The argument was based on 
random motions and also on the modification to the Uncertainity Principle.
We will now consider the problem from a different point of view,
which not only reconfirms the above result, but also enables an elegant
extension to the case of the entire universe itself.
Let us consider an array of $N$ particles, spaced a distance $\Delta x$
apart, which behave like oscillators, that is as if they were connected by
springs. We then have \cite{r2,r3}
\begin{equation}
r  = \sqrt{N \Delta x^2}\label{e1d}
\end{equation}
\begin{equation}
ka^2 \equiv k \Delta x^2 = \frac{1}{2}  k_B T\label{e2d}
\end{equation}
where $k_B$ is the Boltzmann constant, $T$ the temperature, $r$ the extent  and $k$ is the 
spring constant given by
\begin{equation}
\omega_0^2 = \frac{k}{m}\label{e3d}
\end{equation}
\begin{equation}
\omega = \left(\frac{k}{m}a^2\right)^{\frac{1}{2}} \frac{1}{r} = \omega_0
\frac{a}{r}\label{e4d}
\end{equation}
We now identify the particles with Planck masses, set $\Delta x \equiv a = 
l_P$, the Planck length. It may be immediately observed that use of 
(\ref{e3d}) and (\ref{e2d}) gives $k_B T \sim m_P c^2$, which ofcourse agrees 
with the temperature of a black hole of Planck mass. Indeed, as noted, Rosen had shown that a Planck mass particle at the Planck scale  can be considered to be a
universe in itself. We also use the fact alluded to that  a typical elementary particle
like the pion can be considered to be the result of $n \sim 10^{40}$ Planck
masses. Using this in (\ref{e1d}), we get $r \sim l$, the pion
Compton wavelength as required. Further, in this latter case, using (48) and the fact that $N = n \sim 10^{40}$, and (\ref{e2d}),i.e. $k_BT = kl^2/N$ and  (\ref{e3d}) and
(\ref{e4d}), we get for a pion, remembering that $m^2_P/n = m^2,$ 
$$k_ B T = \frac{m^3 c^4 l^2}{\hbar^2} = mc^2,$$
which of course is the well known formula for the Hagedorn temperature for
elementary particles like pions. In other words, this confirms the conclusions
in the previous section, that we can treat an elementary particle as a series of some
$10^{40}$ Planck mass oscillators. However it must be observed from 
(\ref{e2d}) and (\ref{e3d}), that while the Planck mass gives the highest
energy state, an elementary particle like the pion is in the lowest energy
state. This explains why we encounter elementary particles, rather than
Planck mass particles in nature. Infact as already noted \cite{cu}, a Planck
mass particle decays via the Bekenstein radiation within a Planck time
$\sim 10^{-42}secs$. On the other hand, the lifetime of an elementary particle
would be very much higher.\\
In any case the efficacy of our above oscillator model can be seen by the fact that we recover correctly the masses and Compton scales in the order of magnitude sense and also get the correct Bekenstein and Hagedorn formulas as seen above, and get the correct estimate of the mass of the universe itself, as will be seen below.\\
Using the fact that the universe consists of $N \sim 10^{80}$ elementary
particles like the pions, the question is, can we think of the universe as
a collection of $n N \, \mbox{or}\, 10^{120}$ Planck mass oscillators? This is what we will now
show. Infact if we use equation (\ref{e1d}) with
$$\bar N \sim 10^{120},$$
we can see that the extent $r \sim 10^{28}cms$ which is of the order of the diameter of the
universe itself. Next using (\ref{e4d}) we get
\begin{equation}
\hbar \omega_0^{(min)} \langle \frac{l_P}{10^{28}} \rangle^{-1} \approx m_P c^2 \times 10^{60} \approx Mc^2\label{e5d}
\end{equation}
which gives the correct mass $M$, of the universe which in contrast to the earlier pion case, is the highest energy state while the Planck oscillators individually are this time the lowest in this description. In other words the universe
itself can be considered to be described in terms of normal modes of Planck scale oscillators.\\ 
The above gives a rationale for the figure $10^{120}$ Planck oscillators which is derived from the observed $10^{80}$ elementary particles in the universe and considerations in Section 2.

\end{document}